**Title:**

From Cellular Characteristics to Disease Diagnosis: Uncovering Phenotypes with Supercells

**Authors and Affiliations:**

Julián Candia[1,2,3], Ryan Maunu[1], Meghan Driscoll[1], Angélique Biancotto[4,5], Pradeep Dagur[5], J Philip McCoy Jr[4,5], H Nida Sen[6], Lai Wei[6], Amos Maritan[7], Kan Cao[8], Robert B Nussenblatt[6], Jayanth R Banavar[1], Wolfgang Losert[1]

[1]Department of Physics, University of Maryland, College Park, MD 20742, USA
[2]School of Medicine, University of Maryland, Baltimore, MD 21201, USA
[3]IFLYSIB and CONICET, University of La Plata, 1900 La Plata, Argentina
[4]Center for Human Immunology, Autoimmunity and Inflammation, National Institutes of Health, Bethesda, MD 20892, USA
[5]Hematology Branch, National Heart Lung and Blood Institute, National Institutes of Health, Bethesda, MD 20892, USA
[6]Laboratory of Immunology, National Eye Institute, National Institutes of Health, Bethesda, MD 20892, USA
[7]Dipartimento di Fisica "G. Galilei," Università di Padova, Consorzio Nazionale Interuniversitario per le Scienze Fisiche della Materia and Istituto Nazionale di Fisica Nucleare, via Marzolo 8, 35131 Padua, Italy
[8]Department of Cell Biology and Molecular Genetics, University of Maryland, College Park, MD 20742, USA

**Corresponding Author:**

Julián Candia
Address: A.V. Williams (Bldg # 115), Room 3359, University of Maryland, College Park, MD 20742, USA
Phone: (574) 386-1896
Email: candia@umd.edu

**Abstract:**

Cell heterogeneity and the inherent complexity due to the interplay of multiple molecular processes within the cell pose difficult challenges for current single-cell biology. We introduce an approach that identifies a disease phenotype from multiparameter single-cell measurements, which is based on the concept of "supercell statistics", a single-cell-based averaging procedure followed by a machine learning classification scheme. We are able to assess the optimal tradeoff between the number of single cells averaged and the number of measurements needed to capture phenotypic differences between healthy and diseased patients, as well as between different diseases that are difficult to diagnose otherwise. We apply our approach to two kinds of single-cell datasets, addressing the diagnosis of a premature aging disorder using images of cell nuclei, as well as the phenotypes of two non-infectious uveitides (the ocular manifestations of Behçet's disease and sarcoidosis) based on multicolor flow cytometry. In the former case, one nuclear shape measurement taken over a group of 30 cells is sufficient to classify samples as healthy or diseased, in agreement with usual laboratory practice. In the latter, our method is able to identify a minimal set of 5 markers that accurately predict Behçet's disease and sarcoidosis. This is the first time that a quantitative phenotypic distinction between these two diseases has been achieved. To obtain this clear phenotypic signature, about one hundred $CD8^+$ T cells need to be measured. Although the molecular markers identified have been reported to be important players in autoimmune disorders, this is the first report pointing out that $CD8^+$ T cells can be used to distinguish two systemic inflammatory diseases. Beyond these specific cases, the approach proposed here is applicable to datasets generated by other kinds of state-of-the-art and forthcoming single-cell technologies, such as multidimensional mass cytometry, single-cell gene expression, and single-cell full genome sequencing techniques.

**Author Summary:**

The behavior of organisms is based on the concerted action occurring on an astonishing range of scales from the molecular to the organismal level. Molecular properties control the function of a cell, while cell ensembles form tissues and organs, which work together as an organism. In order to understand and characterize the molecular nature of the emergent properties of a cell, it is essential that multiple components of the cell are measured simultaneously in the same cell. Similarly, multiple cells must be measured in order to understand health and disease in the organism. In this work, we develop an approach that is able to determine how many cells, how many measurements per cell, and which measurements are needed to reliably diagnose disease. We apply this method to two different problems: the diagnosis of a premature aging disorder using images of cell nuclei, and the distinction between two similar autoimmune eye diseases using stained cells from patients' blood samples. Our findings shed new light on the role of specific kinds of immune system cells in systemic inflammatory diseases and may lead to improved diagnosis and treatment.

**Introduction**

In the life sciences, there is now a wealth of quantitative information from simultaneous measurements on many proteins and genes, from small tissue samples down to a single cell at a time [1-6]. Likewise, bioimaging is following a similar trend through multicolor fluorescent imaging and the emerging ability to carry out spatially resolved vibrational spectroscopy of living cells in close to real-time [7,8]. These groundbreaking technologies have resulted in a plethora of information for single cells, which can be represented as points in a high-dimensional space. Here we show how one can tease out the essential information from such high-dimensional data in order to diagnose human diseases and understand their molecular origins.

Our approach tackles two interlinked challenges inherent to high-dimensional, single-cell information. First, single-cell measurements exhibit vast heterogeneity in the behavior of individual cells: even a simple bell-shaped distribution can contain subpopulations enriched for biologically distinct functions. For instance, subpopulations of clonally derived hematopoietic progenitor cells with low or high expression of the stem cell marker Sca-1 were observed to be in dramatically different transcriptional states and to give rise to different blood cell lineages [9]. Second, cell phenotypes are emergent products of multiple molecular actions: the phenotype of a tissue or organism often requires not only multiple cells, but also multiple attributes at the cellular level, which makes bridging scales from molecular and cellular level information to disease diagnosis a challenging, oftentimes elusive goal [10].

Here we present a new approach to analyze high-dimensional single-cell information, and apply it to two representative datasets. We address the diagnosis of progeria, a premature aging disorder [11], where single-cell data are obtained by an automated nuclear shape analysis from hundreds of healthy and diseased cells. We also develop a multiparameter phenotype in order to distinguish two sight threatening non-infectious uveitides, the ocular manifestations of Behçet's disease and sarcoidosis, based on multicolor flow cytometry information on tens of proteins from fresh blood patient samples. Our emphasis is to assess the optimal tradeoff between the number of single cells averaged and the number of measurements needed to capture phenotypic difference. The number of available cells may be a key limiting factor when target cell subpopulations are extremely small (e.g. hematopoietic stem cells from bone marrow or blood samples) or when the experimental techniques are not easily scalable (e.g. single-cell imaging and single-cell gene expression).

In the next Section, we describe some common approaches to analyze multidimensional single-cell datasets, we show their shortcomings due to cell heterogeneity and the inherent multidimensional nature implied in a complex phenotype, and we apply our approach to the two specific cases mentioned above. In the following Section, we provide a summary and a discussion of our findings.

**Results**

A commonly used method to visualize and analyze multidimensional single cell information is through sequential selection of subtypes of cells based on simple thresholds, applied to one or two parameters at a time [12]. This procedure is generally represented as a sequence of two-dimensional plots, where one attribute is plotted against another one. This method works extremely well when simple thresholds for just a few parameters lead to reliable phenotypes. However, for complex diseases such as Behçet's and sarcoidosis, even the best choice of parameters is not enough to identify a phenotype. A representative example is shown in Fig. 1A(i): $CD8^+$ T cells have very similar combinations of IL22 and CD3 levels in both Behçet's disease and sarcoidosis, even though – as we will show below – these parameters play a key role in distinguishing between the two diseases. Similarly, highly overlapping populations are observed for other cell types we investigated (e.g. $CD4^+$ T cells) and other pairs of markers studied. This indicates that the distinction between Behçet's disease and sarcoidosis can only be discerned using a combination of more than two parameters, and thus is difficult to visualize and detect with established approaches.

Going beyond two parameters, some mathematical tools are able to reduce the dimensionality of high-dimensional data [13,14]. Singular value decomposition is a simple, yet powerful technique for generating low dimensional representations [13]. However, the optimal axes selected by such a method are not designed to distinguish between health and disease, or help diagnose the disease. This is evident in Fig. 1A(ii), where the two top eigenmodes from a singular-value decomposition analysis of 16-dimensional data are plotted for the same $CD8^+$ T cell subpopulation, showing again a large overlap between the two diseases.

Even in cases where a single parameter can be established as a suitable phenotype, cell-to-cell heterogeneity presents a challenge. For example, in Hutchinson-Gilford progeria syndrome (HGPS), a rare genetic disease of accelerated aging, the number of "blebs" or localized protrusions visible in a cell's nucleus is an established cellular marker of HGPS [15]. However, that does not imply that a single cell showing blebs indicates HGPS. Instead, as shown by Fig. 1B(i)-(iv), blebbed and non-blebbed nuclei are observed both within healthy and diseased cell lines. On average, nevertheless, blebbing is a reliable phenotype, as illustrated in Fig. 1B(v)-(vi). This raises the question: can one simply measure other aspects of the nuclear shape with additional metrics to establish a disease phenotype from a single cell, or does cell heterogeneity require us to investigate the properties of cell ensembles for a reliable diagnosis? The tradeoff between multidimensional measurements and the number of cells needed to achieve a desired confidence level of prediction certainly requires an unbiased, fully quantitative, and mathematically robust method.

Here we introduce and apply an approach to develop a disease phenotype from multiparameter single-cell measurements. Our approach uses simple machine learning methods to determine what combination of parameters can serve as an indicator of disease, and how many parameters are needed to diagnose a disease. While machine learning of disease diagnostics is not new, it often fails when applied at the single-cell level due to the heterogeneity of cells. It also fails when average quantities are measured if the number of patients is not large enough for a machine learning approach. The simple additional step of averaging over a small number of cells - here tens to hundreds of cells – and varying that number allows us to optimize our ability to detect a disease phenotype. This procedure smoothes out single cell heterogeneity and, at the same time, minimizes the loss of information due to averaging. For machine learning purposes, each patient is still represented by a point cloud in parameter space, but now each point represents a group of cells, rather than an individual cell.

Recently, several groups have developed computational methods for identifying cell populations in multidimensional flow cytometry data. Their goals are two-fold: on the one hand, to determine whether automated algorithms can reproduce expert manual gating; on the other hand, to determine whether analysis pipelines can identify characteristics that correlate with external variables such as clinical outcome. In the latter case, flow cytometry data is transformed into class-labeled vectors in instance space by a variety of methods such as binning of 2D and 3D measurement histograms, Gaussian mixtures, 1D and sequential gating schemes, and cell clustering using k-means and other high-dimensional clustering techniques [16-23]. A detailed description and comparative assessment of the performance of different approaches has been recently reported [24]. Within this context, it is important to point out that the method proposed in our work addresses the problem of phenotypic classification when single cells are highly heterogeneous and when the number of cells available may be rather small (just a few tens or hundreds, as opposed to typical flow cytometry experiments in which the number of measured cells is one or several orders of magnitude larger). We will demonstrate that our method is generally applicable to different kinds of multidimensional single-cell data and one of our examples is on flow- cytometry-based phenotypes. However, the key contribution is the development of a framework that provides a quantitative assessment of the critical sample size and number of simultaneous single-cell measurements needed to identify a phenotype with strong predictive power. State-of-the-art single-cell genomics and single-cell imaging technologies are examples in which the number of measured  single cells is critically small, and where flow cytometry data analysis methods that rely on high-dimensional clustering procedures, Gaussian mixture approximations, etc may be expected to fail.

We will tackle the tradeoff between the number of parameters and the number of cells needed first on the example of HGPS - the mathematics are the same for any multidimensional single-cell dataset. A complete approach would entail the study of the distribution of the individual measurement vectors. Our results demonstrate, a posteriori, that simple averages suffice for carrying out the calculations successfully. We define a "supercell of size N" as the average of the individual measurement vectors of N randomly selected cells. By repeatedly taking different random subsets of N cells from the original datasets, we build "supercell samples" and we are thus able to compute "supercell statistics". This procedure is illustrated in Fig. 2A(i)-(ii), where we select one shape parameter (namely, the number of invaginations of the nuclear boundary) and compute the probability density distributions for healthy and diseased cell lines. In Fig. 2A(i), the distributions for single cells are highly overlapping, reflecting the fact that, based on individual cells, one is not able to distinguish healthy cells from diseased ones (SI Dataset S1). After applying the cell averaging procedure (using N=30 randomly selected cells to generate each "supercell"), we obtain distributions without any significant overlap between healthy and diseased samples, as shown by Fig. 2A(ii). The supercell size N=30 has been chosen because it represents the smallest size that provides a full separation between healthy and diseased samples, regardless of the number of parameters used (see discussion below).

The removal of distribution overlaps is a manifestation of the central limit theorem (CLT) of probability theory [25-27]. The CLT states that, given a set of n independent random variables associated with arbitrary probability distributions with finite mean $\mu_i$ and variance $\sigma_i^2$ (for i=1,2,...,n), their average is a random variable whose asymptotic cumulative distribution function approaches a normal distribution with mean $\mu = \sum \mu_i/n$ and variance $\sigma^2 = (\sum \sigma_i^2/n)/n$. As a consequence, distributions of supercells of size N are expected to become narrower by a factor of $\sim 1/\sqrt{N}$. For instance, comparing Fig. 2A(i) with Fig. 2A(ii), we observe that the width of the latter is approximately smaller by a factor of $\sim 1/\sqrt{30} \approx 0.2$. Another consequence of the CLT is that the shape of supercell distributions becomes closer to Gaussian as N is increased. It should be pointed out that the supercell framework does not rely on a priori assumptions regarding the shape of the measurement distributions. On the contrary, it incorporates all features of the original distributions, thus naturally dealing with issues such as skewed distributions with regions that could be ambiguously attributed to outliers or to poorly resolved subpopulations. However, if the measurement distributions are distinctly multimodal due to well-defined cell subpopulations, then the ability to predict reliable phenotypes might be compromised. In such a scenario, robust phenotyping might first require the identification of different cell subpopulations followed by the application of the supercell framework separately to each of them. This procedure is discussed below in the context of distinguishing healthy individuals from patients with two non-infectious uveitides by using either all cells from peripheral blood samples, or different T cell subpopulations (see Fig. 3).

After cell averaging, machine learning allows us to learn what combination of parameters best distinguishes healthy from diseased cells. In order to avoid overfitting and also to obtain a straightforward interpretation of the machine-learned parameters in terms of the original measurements, we used a support vector machine with a linear kernel, which is equivalent to the machine learning method known as the perceptron [28,29]. Healthy and HGPS nuclear shapes were characterized by 12 parameters including eccentricity, number of invaginations, minor/major axis length, mean and standard deviation of the curvature, and perimeter. Moreover, the concentration of lamin A/C (measured based on the fluorescence signals of lamin A/C) was represented through 3 additional parameters for each nucleus. However, for single cells, even with these 15 parameters, the distinction between individual cells from healthy and diseased cell lines is not learnable. Fig. 2B(i) shows the distance from each cell to the perceptron boundary, where positive (negative) distances correspond to the boundary side identified with the healthy (diseased) class. We observe that some cells from the healthy cell lines are classified as diseased, and vice versa. Instead, machine learning applied to the supercell samples works with 100% accuracy, as displayed in Fig. 2B(ii).

The questions arise, then, which and how many parameters are needed to achieve a classification of desired accuracy, and how many cells need to be averaged into a "supercell". Fig. 2C(i) shows the perceptron amplitudes (i.e., the components of the vector normal to the boundary hyperplane) for each of the 15 parameters. A positive sign indicates that a given parameter is higher in healthy cells relative to diseased cells, while its absolute value is a measure of its overall significance (relative to the other parameters) in separating healthy cells from diseased ones. Therefore, we can rank-order the 15 parameters from most to least relevant according to their decreasing amplitudes (in absolute values), and learn using just the top M parameters from the rank-ordered list. While this rank ordering is independent of supercell size for large supercells, it it is very different from the rank ordering for single cells (if the single cell measurements are strongly overlapping). Indeed, sizable fluctuations are observed in the single-cell and small-supercell regime (up to supercells of size ~10) followed by a stable rank-order for larger supercell sizes. The fraction of cells correctly classified by the machine learning process as a function of the supercell size is shown in Fig. 2C(ii). The different curves represent different numbers of parameters (M). As expected, the classification accuracy increases with both M and the supercell size. While a single cell is not sufficient for classification, a single parameter (the number of invaginations) is sufficient for correct classification of HGPS. Indeed, this is consistent with the standard approach to assess the disease states of HGPS based on visual analysis (i.e. the detection of "blebs") and indicates that the invaginations are the most distinguishing features of blebs [30-32].

In our second example, we apply our technique first to the simpler problem of distinguishing healthy individuals from patients with two non-infectious uveitides, and then to the formidable challenge of distinguishing Behçet's disease from sarcoidosis. Recent work has reported progress in the ability to pinpoint molecular indicators for inflammatory immune diseases, where larger-than-normal levels of a novel subset of effector memory $CD4^+$ T lymphocytes expressing the endothelial adhesion molecule CD146 have been observed in sarcoidosis, Behçet's, and Crohn's disease [33]. However, while patients can be diagnosed with Behçet's disease or sarcoidosis based on the concurrent observation of a number of clinical indicators, molecular signatures unique to these diseases have not been found. Our analysis of a molecular phenotype uses flow cytometry experiments, in which 14 molecular markers previously reported on human $CD4^+$ and/or $CD8^+$ T cells were measured for each cell; additionally, forward- (FSC) and side-scattering (SSC) measures were also taken on each cell. Thus, a total of 16 simultaneous measurements were performed on each cell from patients' peripheral blood, with about one million cells measured per patient. From a cohort of 22 patients, 7 were diagnosed with sarcoidosis, 6 with Behçet's disease, 1 with retinal vasculitis, while the remaining 8 were healthy controls. We start with large supercells to assess whether molecular phenotyping is possible at all to distinguish sarcoidosis and Behçet's disease. We represent each patient sample with 100 supercells, where each supercell was obtained from averages over 500 randomly chosen cells. We carry out separate analyses for the distinction between healthy and diseased patients (Fig. 3(a)-(c)), and for the separation between the two diseases sarcoidosis and Behçet's (Fig. 3 (d)-(f)). Furthermore, we perform separate analyses for all cells (SI Dataset S2), for CD4$^+$ T cells (that can be isolated using standard gating procedures based on the sequence viability$^-$/CD3$^+$/CD4$^+$/CD8$^-$) (SI Dataset S3) and for CD8$^+$ T cells (similarly identified according to viability$^-$/CD3$^+$/CD8$^+$/CD4$^-$) (SI Dataset S4).

Because we have a larger number of patients than we did for HGPS, we can directly assess the predictive power of our approach to correctly diagnose a new patient. We tested the predictive power of our learnt patterns using a standard data-resampling method, namely the so-called jackknife procedure: leaving out one patient at a time, one learns with the remaining data and makes a prediction on the test patient [34]. In that way, one can determine the percentage of correct and failed predictions. Since each patient is represented by a cloud of 100 supercells, it may happen that the perceptron boundary cuts across the test patient's supercell cloud. We set a threshold of 95% in order to make a prediction: e.g. if the supercell cloud is more than 95% consistent with sarcoidosis, we classify the patient as having sarcoidosis. If the supercell cloud falls on the boundary between diagnoses (i.e. with less that 95% of the supercells on either side of the perceptron boundary), we leave the test patient unclassified. Naturally, setting the prediction threshold to lower values leads to less unclassified patients, but tends to increase the number of failed predictions; in contrast, increasing the threshold to higher values leads to a more conservative approach, where the number of failed predictions is smaller at the expense of a larger number of unclassified patients. By changing the prediction threshold values over the range between 80% and 100%, the observed variations of the predicted outcome were below 10% of the cohort; the method is thus largely insensitive to the choice of the threshold parameter.

By learning using all available measures, we are able to rank-order the importance of the measures based on the perceptron amplitudes. The ten most important measures and corresponding amplitudes are listed in Fig. 3(a)-(f). The percentage of patients correctly predicted (green), unclassified (blue), and incorrectly predicted (red) are shown as a function of the number of rank-ordered measures used. The outcomes depend strongly on the type of cells used: for the "healthy vs diseased" case, no incorrect predictions are made using all cells and just the top two measures, namely viability and CD197 (Fig. 3(a)). The predictions are even stronger if using only CD4$^+$ T cells, since the top marker (CD27) is sufficient by itself to correctly classify all healthy patients (with high frequency of CD4$^+$CD27$^+$ T cells in their peripheral blood) and all diseased patients (with low frequency of CD4$^+$CD27$^+$ T cells in their peripheral blood) in the cohort (Fig. 3(b)). In contrast, failed predictions are seen for the case of CD8$^+$ T cells, irrespective of the number of measures used (Fig. 3(c)). Previous reports have suggested that CD4$^+$CD27$^+$ T cells represent the majority of natural regulatory T cells in human peripheral blood [35]. Thus, our results indicate that patients with either Behcet's disease or Sarcoidosis have low frequency of peripheral natural regulatory T cells and, therefore, potentially compromised immunoregulatory functions during inflammatory responses.

In order to separate Behçet's disease and sarcoidosis, predictions based on all cells are very poor (Fig. 3(d)), better for CD4$^+$ T cells (Fig. 3(e)) and best for CD8$^+$ T cells (Fig. 3(f)), for which no failed predictions are made when five or more measures are used. This result indicates that the top measures listed in Fig. 3(f) may be used as molecular phenotypes that distinguish the two diseases. This is, to the best of our knowledge, the first report pointing out that CD8$^+$ T cells can be used to distinguish two systemic inflammatory diseases. Moreover, it is interesting to note that, in distinguishing between patients with ocular inflammation and controls without it, the CD4 marker was an important feature, while for distinguishing between the ocular manifestations of two systemic disorders, the CD8 cell marker was superior.

Given our success in demonstrating the power of molecular phenotyping to distinguish the diseases, we turn now to the analysis of the balance between the number of cells we need to average, and the number of molecular markers we need to measure. For the "healthy vs diseased" case using CD4$^+$ T cells, the percentage of correctly classified supercells is shown in Fig. 3(g) as a function of the supercell size and the number of measures used. Note that for single cells, the classification performance is very poor even using many measures, but averaging over more than ten cells is sufficient for reliable classification if a large number of measures is used. In contrast, just one measured marker is sufficient provided that we average over 100+ cells. This fact is underscored in Fig. 3(h), where the intensity distribution for supercells of size 500 are shown separately for the healthy and the diseased patients, using just the top marker (CD27). The dashed line indicates the marker intensity threshold that allows a complete separation of the two classes of supercells. The "sarcoidosis vs Behçet's" classification is further studied in Fig. 3(i) for CD8$^+$ T cells, where the percentage of correctly classified supercells is shown as a function of the supercell size and the number of measures used. We find that slightly less than 100 cells are sufficient for reliable classification, as long as the top five markers are measured. Increasing the number of markers or averaging over more cells does not strongly change the reliability of the classification. Finally, the ability to classify Behçet's disease vs sarcoidosis when using the top 5 markers is visualized in a new way in Fig. 3(j). The visualization is derived from the identification of patterns in two-dimensional parameter space (Fig. 1A(i)), which has proven to be a tremendously successful tool for the analysis of low-dimensional data in flow cytometry. The combined approach of cell averaging into supercells, followed by machine learning, allows us to find the correct linear combinations of markers needed to fully separate the two diseases (Fig. 3(j)). In geometrical terms, we learned that only 5 dimensions (out of the original 16) are needed; moreover, we determined the preferred direction that maximizes the gradient between the "sarcoidosis class" and the "Behçet's disease class". This optimal class separation was achieved by means of an unbiased, mathematically robust method: no additional biological information was needed to proceed from Fig. 1A(i) to Fig. 3(j).

## Discussion

We present a simple approach to quantify disease phenotypes based on single cell measurements with multiple parameters measured on each cell. For our study of autoimmune diseases, we measure 16 parameters for millions of cells with flow cytometry, and use this information to find a molecular phenotype of Behçet's disease. We also measure 15 parameters from hundreds of fluorescence images obtained via microscopy, and use this information to automate classification of HGPS. Our data span many more dimensions than the traditional two parameters used for visually-aided cell classification (Fig. 1A). We use machine learning, which allows for a reproducible, objective, and automated approach to find the optimal boundary between two high-dimensional classes of data points. The question we tackle is straightforward: do we obtain more information about a disease by the analysis of more cells, or by measuring more parameters on each cell?

The key to our novel approach is to introduce variable size cell groups ("supercells"), with the group size as an explicit parameter that we vary systematically. This reveals the number of cells that need to be grouped in order to obtain a robust disease classification. We also determine to what degree adding parameters reduces the number of cells needed to determine a phenotype.

Our approach to separate cell groups relies on a machine learning classification method. It is tailored specifically to determine the most useful combination of parameters to distinguish among all cells or cell groups rather than finding the optimal low-dimensional representation, as in singular value decomposition or principal component analysis. This procedure is schematically illustrated in Fig. 4, where synthetic 2D datasets were generated to represent patient samples classified in two categories: 4 samples correspond to "Class A", while the remaining 3 samples are labeled "Class B". At the single-cell level, the data are non-separable due to cell heterogeneity (Fig. 4(a)). A machine learning classifier such as support vector machines with a linear kernel (Fig. 4(b)) can be implemented in order to find the optimal decision boundary between the two classes; however, this method requires the data to be separable. More sophisticated variants, such as soft margin classifiers and nonlinear classifiers can be designed to learn from non-separable data, even in strongly overlapping cases such as those usually encountered in single-cell datasets (see e.g. Fig. 1A(i)). However, our analysis shows that the optimal "boundary" inferred from overlapping distributions is distinct from the boundary obtained from supercells which actually has predictive power for phenotyping: the weight of parameters is very different for single cells and supercells whose distribution is well separated. In order to avoid overlapping patient samples, one could characterize each of them by the moments of the cell multivariate distributions, the simplest example being the sample means (Fig. 4(c)). This approach, however, lacks robustness: the decision boundaries are very sensitive to nearby datapoints, in particular to the support vectors that determine the classification hyperplanes, thus leading to failed predictions. Supercell distributions are built by averaging over groups of single cells. By applying machine learning on supercell samples, a robust class separation is achieved (Fig. 4(d)).

In HGPS, our approach confirms the current practice that the number of invaginations (or mean negative curvature) is the most valuable nuclear metrics for phenotyping the disease using nuclear images. Importantly, we find that when analyzing 30 cells or more, a robust phenotype can be obtained simply based on the invaginations of each cell, and a more in depth analysis of additional nuclear shape metrics does not significantly reduce the number of cells needed. Our findings provide a principle guideline of the minimal cell numbers used in future disease assessments and high-throughput drug screenings of age-related diseases, in which abnormal nuclear shape is considered a hallmark phenotype. This information is of extreme importance in a rare disease like HGPS with very limited availability of patient samples.

In our second example, we apply our technique to distinguish healthy individuals from patients with two non-infectious uveitides, and among those patients we distinguish between Behçet's disease and sarcoidosis. In order to distinguish healthy from disease phenotypes, we found that within the $CD4^+$ T cell subpopulation, just one marker was enough (Fig. 3(b)). Indeed, CD27 appears consistently overexpressed in healthy samples (Fig. 3(h)). The ability to predict healthy and diseased phenotypes based on $CD4^+$ T (super)cells is resilient under the removal of the top markers: even by removing the top 7 markers from the list, we are still able to classify patients as healthy or diseased with no failures. In contrast, $CD8^+$ T cells do not have a clear distinction between healthy and diseased conditions, even using all markers available from the flow cytometry experiment (Fig.3 (c)). However, by focusing specifically on sarcoidosis and Behçet's disease, we demonstrate a robust means of predicting a patient's diagnosis based on 5 optimally chosen markers using $CD8^+$ T (super)cells (Fig. 3(f)). If the top marker (IL22) is removed from the list, incorrect predictions are observed even using all remaining markers; therefore, phenotyping sarcoidosis vs Behçet's is inherently high-dimensional (since it requires at least 5 markers to be accurate) and also very specific to those markers. An important evaluation for the future will be to evaluate the efficacy of these markers in patients with these two systemic disorders who do not have ocular complications of their disease, i.e. whether these findings are specific to the ocular disorder, or a reflection of the systemic disorder itself. By using a precise linear combination of IL22, CD3, viability, CD8 and CD62L, we are able to separate the two diseases successfully based on molecular markers (Fig. 3(j)). Averages of hundreds of cells are required for this phenotyping, and increasing the number of measured parameters does not reduce the number of cells required. The molecular markers used have been reported to be important players in autoimmune disorders. Yang et al. [36] reported an increased number of Th22 cells and increased serum IL-22 levels in patients with lupus skin disease, but a decrease in patients with lupus nephritis. CD62L has been reported to be associated with $CD4^+CD25^{bright}FOXP3^+$ cells in bullous pemphigoid patients [37]. Finally, expanded clones of $CD8^+$ T lymphocytes are present in the lesions of multiple sclerosis [38]. Based on the observations from the analyses presented here, our evaluation of $CD8^+$ T cells has permitted us to see CD8-subset differences in this cell type in patients diagnosed with different uveitides.

Our ability to study the tradeoff between measuring more parameters or analyzing more cells, as shown in Figs. 3(g) and 3(i), has far-reaching consequences for a number of emerging technologies that allow for multi-parameter single-cell measurements. For more challenging problems than those considered here, it may become necessary to study the distributions of the measurement vectors of individual cells rather than its principal surrogate of the first moment, and extend the machine learning algorithms to well-chosen non-linear kernels. High-throughput automated microscopy, where thousands of cells are imaged automatically, is quickly becoming the norm, calling for reliable approaches to classify observations and quantify phenotypes. Similarly, while simultaneous (multicolor) measurement of 16 parameters is the current state-of-the-art for flow cytometry, a next generation of high-throughput single-cell analysis tools is emerging that will allow the measurement of more than 50 parameters at comparable high throughput by means of mass cytometry [3,4]. It is now also becoming possible to analyze gene sequences or gene expression levels for individual cells, although the cost of these expensive technologies severely limits the sample size to much fewer cells than flow cytometry [5,6]. Optimizing the tradeoff between measuring more cells or more parameters, as we demonstrate here, should allow us to take full advantage of these powerful and promising next-generation single-cell technologies.

**Methods**

**Ethics Statement:**
This investigation was conducted according to the principles expressed in the Declaration of Helsinki and was approved by institutional review boards at National Eye Institute, National Institutes of Health. The written informed consent was provided by all patients.

For the study of Hutchinson-Gilford progeria syndrome, cultured fibroblasts from two patients (HGADFN164-p15 and HGADFN167-p15) and two healthy individuals (HGADFN090-p15 and HGADFN168-p15) were used. The cells were fed with fresh MEM medium containing 15% FBS and grown at $37^0$C. In order to visualize the nuclei, we performed immunofluorescence staining of the nuclear membrane with a mouse monoclonal antibody raised against lamin A/C. (MAB3211). This antibody has been well characterized in HGPS cells and has also been used in studies of other laminopathies. Fluorescence images of about 600 nuclei per cell line were taken with a Zeiss fluorescence microscope at 400X magnification, as shown in the examples from Figure 1B. Following the procedure from Driscoll et al (15), a custom-written MATLAB program was used to extract nuclear shapes and their properties, such as the number of invaginations, the mean curvature, the standard deviation of the curvature, etc. In addition to 12 shape measurements, we obtained 3 measurements of the intensity of immunofluorescence from lamin A/C associated with each nucleus (the full list of measurements is provided in Figure 2C(i).

For the study of non-infectious uveitides, peripheral blood samples were obtained from a cohort of 22 patients, out of which 7 were diagnosed with sarcoidosis, 6 with Behçet's disease, 1 with retinal vasculitis, while the remaining 8 were healthy controls. 3 different marker panels were studied on each sample, each consisting of 2 scattering measurements (FSC and SSC) plus 14 or 15 cell surface fluorochromes. Some common markers (such as CD3, CD4, CD8, CD27, CD45, and viability) were used on all 3 panels and were checked for consistency. Separate analyses have been performed on each set of markers in order to find the best prediction accuracy. Two marker panels did not lead to accurate sarcoidosis vs Behçet's disease phenotypes; the third one, which led to an accurate phenotype and has been discussed throughout, consisted of FSC, SSC, IL23R, CD196, CD4, viability, CD8, CD27, CD45, IL17A, CD197, CD3, IL22, CD62L, CD161, and TNFA.

Pre-processed datasets are provided as Supporting Information. Multicolor flow cytometry raw datasets are available at the Dryad Repository: http://dx.doi.org/10.5061/dryad.v6st3.

Data analysis was performed using custom-written programs in R and Perl.

**Supporting Information**

**SI Dataset S1:** Nuclear shape and lamin A/C measurements for 2 healthy and 2 HGPS cell lines.
**SI Dataset S2:** Multicolor flow cytometry (all cells) from 22 patients labeled according to disease type or healthy status for a randomized single-cell subsample and for the different supercell sizes used in this paper.
**SI Dataset S3:** Multicolor flow cytometry ($CD4^+$ T cells) from 22 patients labeled according to disease type or healthy status for a randomized single-cell subsample and for the different supercell sizes used in this paper.
**SI Dataset S4:** Multicolor flow cytometry ($CD8^+$ T cells) from 22 patients labeled according to disease type or healthy status for a randomized single-cell subsample and for the different supercell sizes used in this paper.

**Acknowledgments**

The primary fibroblast cell lines (normal1= HGFDFN090, normal2=HGFDFN168, HGPS1=HGADFN164, and HGPS2=HGADFN167) were obtained from the Progeria Research Foundation (K.C.). The cell staining and imaging was performed by the Cao lab (Christina LaDana). We acknowledge Yang Shen for discussions.

**Financial Disclosure**

KC was supported by NIH grant R00AG029761. WL and JC were partially supported by NIH grants R01GM085574 and PHY1205965. JC was partially supported by CONICET (Argentina) and NIH Project Number 5T32CA154274-03 (T32 Training Grant in Cancer Biology). The funders had no role in study design, data collection and analysis, decision to publish, or preparation of the manuscript.

**References**

1. Sachs K, Perez O, Pe'er D, Lauffenburger DA, Nolan GP (2005) Causal Protein-Signaling Networks Derived from Multiparameter Single-Cell Data. Science 308, 523-529.

2. Tang F, Barbacioru C, Wang Y, Nordman E, Lee C, et al. (2009) mRNA-Seq whole-transcriptome analysis of a single cell. Nature Methods 6:377-384.

3. Bendall SC, Simonds EF, Qiu P, Amir ED, Krutzik PO, et al. (2011) Single-Cell Mass Cytometry of Differential Immune and Drug Responses Across a Human Hematopoietic Continuum. Science 332, 687-696.

4. Benoist C, Hacohen N (2011) Flow Cytometry, Amped Up. Science 332:677-678.

5. Sen A, Rothenberg ME, Mukherjee G, Feng N, Kalisky T, et al. (2012) Innate immune response to homologous rotavirus infection in the small intestinal villous epithelium at single-cell resolution. Proc Natl Acad Sci USA 109(50):20667-20672.

6. MacArthur BD, Sevilla A, Lenz M, Müller F-J, Schuldt BM, et al. (2012) Nanog-dependent feedback loops regulate murine embryonic stem cell heterogeneity. Nature Cell Biol 14:1139-1147.

7. Wang H-W, Chai N, Wang P, Hu S, Dou W, et al. (2011) Label-free bond-selective imaging by listening to vibrationally excited molecules. Phys Rev Lett 106:238106.

8. Afonso PV, Janka-Junttila M, Lee YJ, McCann CP, Oliver CM, et al. (2012) LTB4 is a signal-relay molecule during neutrophil chemotaxis. Dev Cell 22(5):1079-1091.

9. Chang HH, Hemberg M, Barahona M, Ingber DE, Huang S (2008) Transcriptome-wide noise controls lineage choice in mammalian progenitor cells. Nature 453:544-547.

10. Altschuler SJ, Wu LF (2010) Cellular Heterogeneity: Do differences make a difference? Cell 141: 559-563.

11. Capell BC, Collins FS (2006) Human laminopathies: nuclei gone genetically awry. Nat Rev Genet 7(12):940-952.

12. Carey JL, McCoy Jr JP, Keren DF (2007) Flow cytometry in clinical diagnosis (4th Ed.). Chicago: American Society for Clinical Pathology Press. 384 p.

13. Holter NS, Mitra M, Maritan A, Cieplak M, Banavar JR, et al. (2000) Dynamic modeling of gene expression data. Proc Natl Acad Sci USA 97:8409-8414.

14. Treiser MD, Yang EH, Gordonov S, Cohen DM, Androulakis IP, et al. (2010) Cytoskeleton-based forecasting of stem cell lineage fates. Proc Natl Acad Sci USA 107:610-615.

15. Driscoll MK, Albanese JL, Xiong Z-M, Mailman M, Losert W, et al. (2012) Automated image analysis of nuclear shape: What can we learn from a prematurely aged cell? Aging 4(2):119-132.

16. Hahne F, LeMeur N, Brinkman RR, Ellis B, Haaland P, et al. (2009) flowCore: a Bioconductor package for high throughput flow cytometry. BMC Bioinformatics 10: 106.

17. Naim I, Sharma G, Datta S, Cavenaugh JS, Wang J-C, et al. (2010) Swift: scalable weighted iterative sampling for flow cytometry clustering. IEEE Acoustics Speech and Signal Processing 509–512.

18. Qian Y, Wei C, Lee FE-H, Campbell J, Halliley J, et al. (2010) Elucidation of seventeen human peripheral blood B-cell subsets and quantification of the tetanus response using a density-based method for the automated identification of cell populations in multidimensional flow cytometry data. Cytometry B Clin. Cytom. 78 (suppl. 1): S69–S82.

19. Aghaeepour N, Jalali A, O'Neill K, Chattopadhyay PK, Roederer M, et al. (2012) RchyOptimyx: Cellular Hierarchy optimization for flow cytometry. Cytometry Part A 81A: 1022-1030.

20. Qiu P (2012) Inferring phenotypic properties from single-cell characteristics. PLoS ONE 7(5): e37038.

21. Ge Y, Sealfon SC (2012) flowPeaks: a fast unsupervised clustering for flow cytometry data via K-means and density peak finding. Bioinformatics 28: 2052–2058.

22. Aghaeepour N, Chattopadhyay PK, Ganesan A, O'Neill K, Zare H, et al. (2012) Early immunologic correlates of HIV protection can be identified from computational analysis of complex multivariate T-cell flow cytometry assays. Bioinformatics 28: 1009–1016.

23. Zare H, Bashashati A, Kridel R, Aghaeepour N, Haffari G, et al. (2012) Automated analysis of multidimensional flow cytometry data improves diagnostic accuracy between mantle cell lymphoma and small lymphocytic lymphoma. Am. J. Clin. Pathol. 137: 75–85.

24. Aghaeepour N, Finak G, The FlowCAP Consortium, The DREAM Consortium, Hoos H, et al. (2013) Critical assessment of automated flow cytometry data analysis techniques. Nature Methods 10(3): 228-238.

25. Kallenberg, O (1997) Foundations of Modern Probability (2nd Ed.). New York: Springer-Verlag. 664 p.

26. Rice, JA (2006) Mathematical Statistics and Data Analysis (3rd Ed.). Belmont: Duxbury Press. 688 p.

27. Jin EY, Reidys CM (2008) Central and local limit theorems for RNA structures. J. Theor. Biol. 250(3):547-559.

28. Cristianini N, Shawe-Taylor J (2000) An Introduction to Support Vector Machines and Other Kernel-based Learning Methods. Cambridge: Cambridge University Press. 204 p.

29. Witten IH, Frank E, Hall MA (2011) Data Mining: Practical Machine Learning Tools and Techniques (3rd Ed.). Burlington: Morgan Kaufmann Publishers. 664 p.

30. Capell BC, Erdos MR, Madigan JP, Fiordalisi JJ, Varga R, et al. (2005) Inhibiting farnesylation of progerin prevents the characteristic nuclear blebbing of Hutchinson-Gilford progeria syndrome. Proc Natl Acad Sci USA 102(36):12879-84.

31. Cao K, Capell BC, Erdos MR, Djabali K, Collins FS (2007) A lamin A protein isoform overexpressed in Hutchinson-Gilford progeria syndrome interferes with mitosis in progeria and normal cells. Proc Natl Acad Sci USA 104(12):4949-54.

32. Cao K, Graziotto JJ, Blair CD, Mazzulli JR, Erdos MR, et al. (2011) Rapamycin reverses cellular phenotypes and enhances mutant protein clearance in Hutchinson-Gilford progeria syndrome cells. Sci Transl Med. 3(89):89ra58.

33. Dagur PK, Biancotto A, Wei L, Sen HN, Yao M, et al. (2011) MCAM-expressing $CD4^+$ T cells in Peripheral Blood Secrete IL-17A and are Significantly Elevated in Inflammatory Autoimmune Diseases. J Autoimmun 37(4):319-327.

34. Shao J and Tu D (1995) The Jackknife and Bootstrap. New York: Springer-Verlag. 517 p.

35. Mack DG, Lanham AM, Palmer BE, Maier LA, Fontenot AP (2009) CD27 expression on $CD4^+$ T cells differentiates effector from regulatory T cell subsets in the lung. J Immunol 182(11):7317-24.

36. Yang XY, Wang HY, Zhao XY, Wang LJ, Lv QH, et al. (2013) Th22, but not Th17 Might be a Good Index to Predict the Tissue Involvement of Systemic Lupus Erythematosus. J Clin Immunol 33(4):767-74.

37. Quaglino P, Antiga E, Comessatti A, Caproni M, Nardò T, et al. (2012) Circulating $CD4^+$ $CD25^{bright}FOXP3^+$ regulatory T-cells are significantly reduced in bullous pemphigoid patients. Arch Dermatol Res 304(8):639-645.

38. Ifergan I, Kebir H, Alvarez JI, Marceau G, Bernard M, et al. (2011) Central nervous system recruitment of effector memory CD8$^+$ T lymphocytes during neuroinflammation is dependent on α4 integrin. Brain 134(Pt 12):3560-3577.

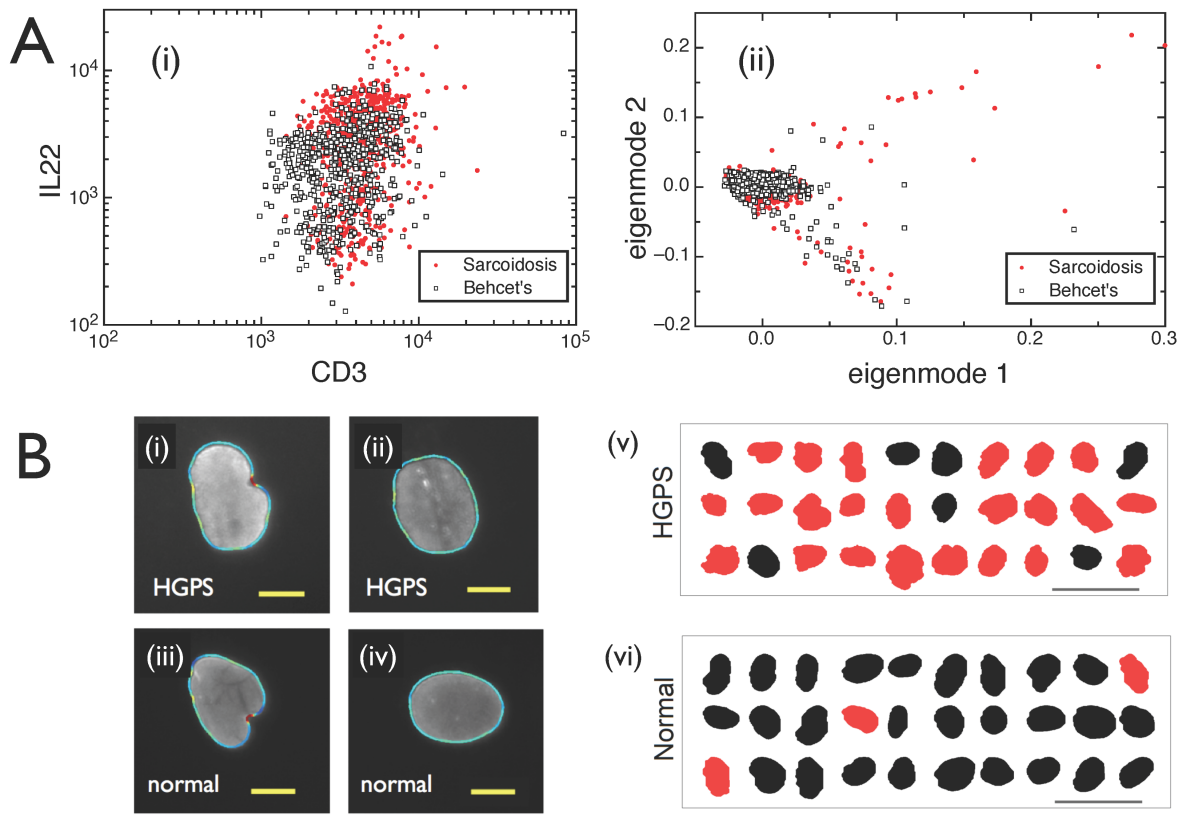

**FIGURE 1:**

**Identifying diseases from heterogeneous single cells.**
**A. Using standard methods of flow cytometry analysis, diseases such as sarcoidosis and Behçet's cannot be separated. (i)** 2D scatter plot using markers CD3 and IL22. **(ii)** 2D Singular-Value decomposition analysis. Figs. A(i)-(ii) show CD8+ T cell subsamples from a cohort of 7 sarcoidosis patients and 6 patients diagnosed with Behçet's disease, but similar overlaps are also observed for other cell types and marker pairs. **B. Cell ensembles carry the signatures of health and disease, despite heterogeneity at the single-cell level. (i)-(iv)** Nuclear shapes of healthy and diseased (HGPS) cells can be classified as either blebbed or non-blebbed. Scale bar: 10$\mu$. Note that it is impossible to tell whether a person has the disease or not based on the analysis of a single cell. **(v)** Classifying nuclei as blebbed (red) or non-blebbed (black) based on just one shape parameter, which is automatically determined via custom image analysis software. Most cells in the ensemble of 30 randomly selected nuclei from a diseased cell line are labeled as blebbed. Scale bar: 50$\mu$. **(vi)** Conversely, analyzing nuclei from a healthy cell line, most cells are labeled as non-blebbed.

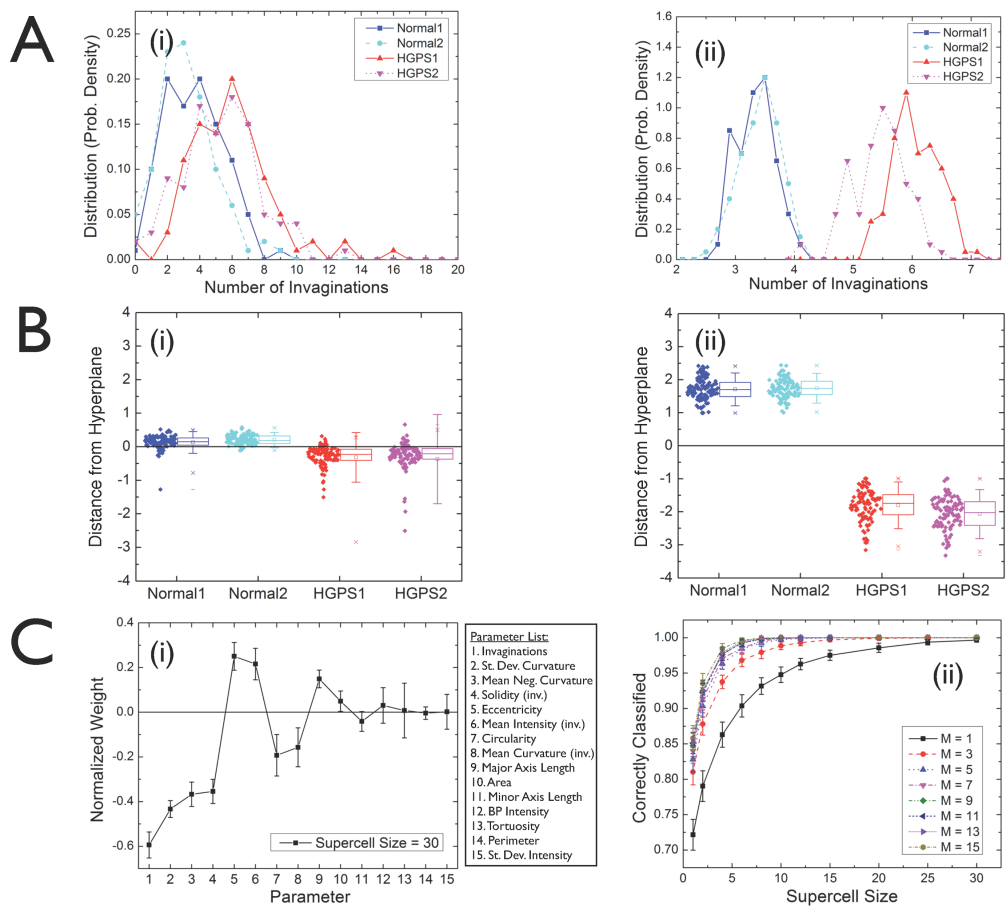

**FIGURE 2:**

**Quantitative multiparameter phenotyping of healthy and HGPS cells through cell averaging ("supercells") and machine learning. A.** Probability density distributions for one shape parameter (number of invaginations of the nuclear boundary) for healthy and diseased cell lines: **(i)** single cells; **(ii)** supercells of size 30. The cell averaging procedure removes the overlap between healthy and diseased cell line distributions. **B.** Distance from the perceptron boundary after machine learning, where positive (negative) distances correspond to the boundary side identified with the healthy (diseased) class: **(i)** single cells; **(ii)** supercells of size 30. Each cell line is shown separately along the horizontal axis. **C. (i)** Perceptron amplitudes: components of the vector normal to the classification hyperplane, each one associated with one of the shape parameters shown in the list. **(ii)** Fraction of cells correctly classified by the machine learning process as a function of the supercell size for a varying number of parameters used, as indicated. The top M measures are selected from the rank-ordered list based on the absolute values of the perceptron amplitudes.

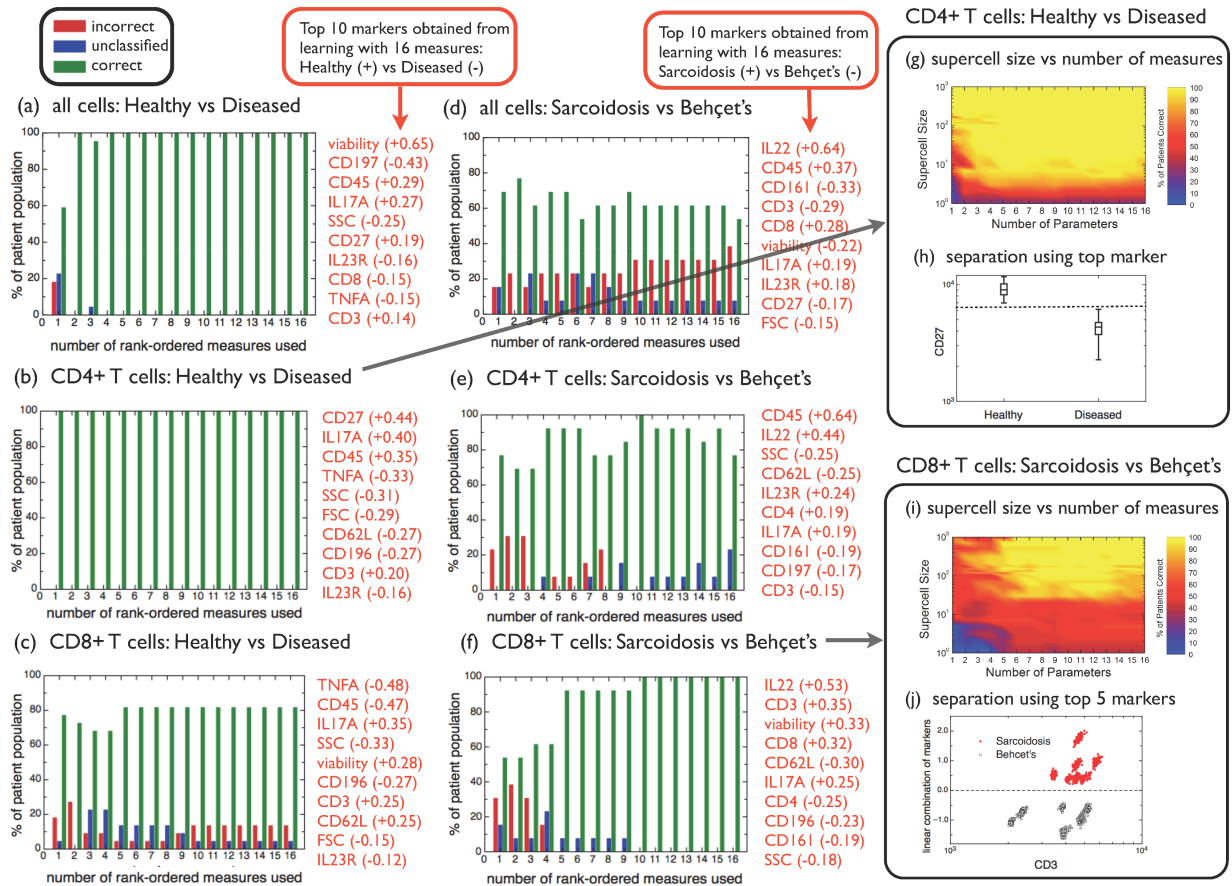

**FIGURE 3:**
**Predictive power of automated phenotyping to distinguish healthy vs diseased, or sarcoidosis vs Behçet's for different cell types and number of markers measured.** **(a)-(f)** A jackknife analysis of patient classification was carried out based on a sample with 100 supercells, where each supercell was obtained from averages over 500 randomly chosen cells. The percentage of patients correctly classified is shown as green bars, the percentage of patients for which a classification is not possible (because less than 95% of supercells fall into either one of the classes) is shown as blue bars, while the percentage of patients incorrectly classified is shown as red bars. The top 10 measures for each case are listed to the right of each plot. Separate analyses have been carried out for all cells, CD4+ T cells and CD8+ T cells, as indicated, as well as for the two binary classification scenarios "healthy vs diseased", and "sarcoidosis vs Behçet's". **(g)** Percentage of supercells correctly classified as healthy or diseased, as a function of the supercell size and the number of measures used, within the CD4+ T cell subpopulation. **(h)** Distribution of the top marker (CD27) for supercells averaged over 500 randomly chosen CD4+ T cells. **(i)** Percentage of supercells correctly classified as sarcoidosis or Behçet's disease, as a function of the supercell size and the number of measures used, within the CD8+ T cell subpopulation. **(j)** Linear combination of the top 5 markers IL22, CD3, viability, CD8 and CD62L, as a function of CD3, for supercells averaged over 500 randomly chosen CD8+ T cells.

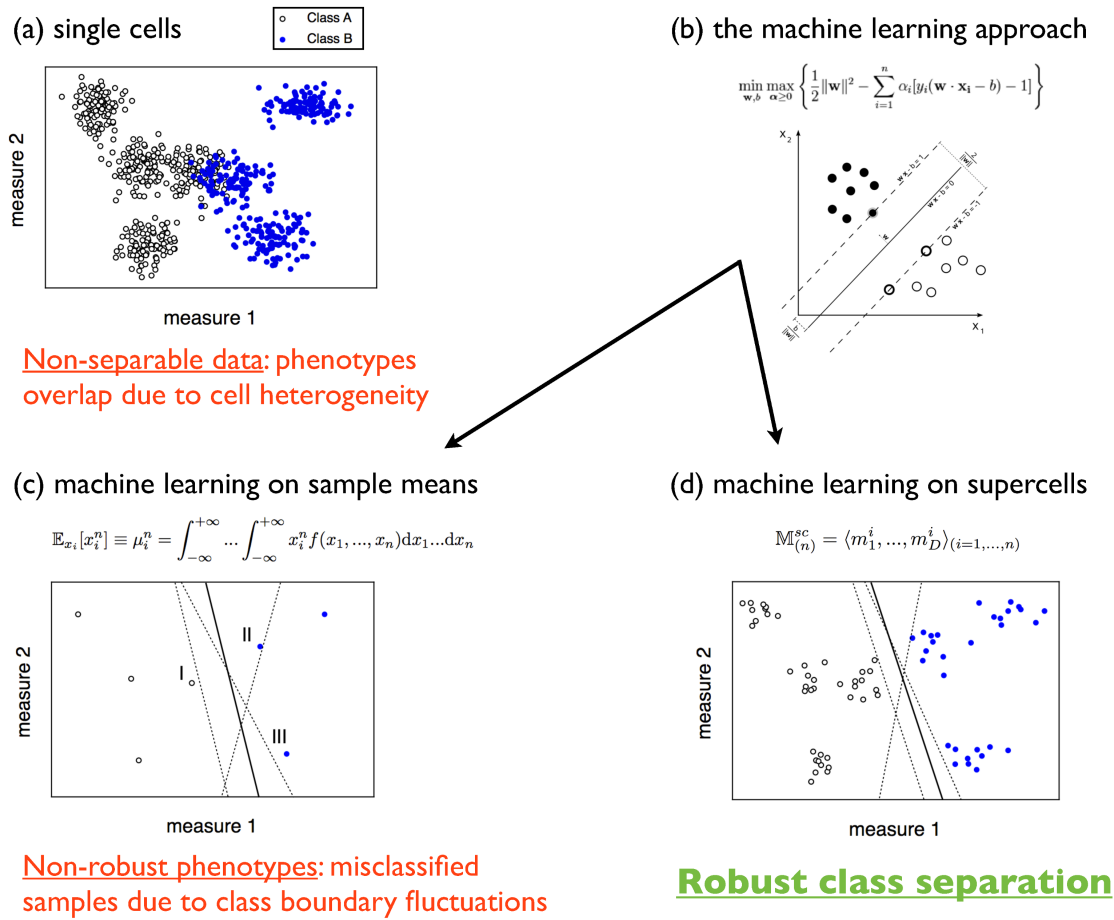

**FIGURE 4:**

**Summary of the supercell approach. (a)** 2D synthetic data representing 7 single-cell patient samples in two categories. Due to cell heterogeneity, different phenotypes overlap and the data are non-separable. **(b)** A machine learning approach such as support vector machines is able to find the optimal decision boundary between two classes of datapoints. However, this method (and variants thereof) fail when the samples are strongly overlapping, as is the usual case for single-cell datasets (recall Fig. 1A(i)). **(c)** Sample means or higher-order moments of the cell multivariate distributions generally lead to poor, non-robust phenotypes. The solid line is the class boundary learnt using all datapoints; by removing either of the support vectors that define this boundary (marked by "I", "II", and "III"), the boundary changes as indicated by the dashed lines, thus leading to jackknife prediction failures. **(d)** Representing patient samples by supercell distributions, class separation becomes robust. Removing patient samples "I", "II", or "III", the decision boundary changes as shown by the dashed lines. Departures from the boundary learnt using all patients (solid line) are less significant and do not cause any jackknife failed predictions.